\documentclass[12pt]{article}
\usepackage[margin=1in]{geometry}
\usepackage{graphicx}
\usepackage{xcolor}
\usepackage{comment}
\usepackage{ulem}
\usepackage{cite}
\usepackage[caption=false]{subfig}
\usepackage{caption}
\usepackage{lineno}

\newcommand{\lineup}{}
\newcommand{\br}{\hline}
\newcommand{\mr}{\hline}
\newcommand{\ack}{\section*{Acknowledgments}}

\begin{document}

\title{Sawtooth suppression by flux pumping on HBT-EP}

\author{
Boting Li$^{1}$, J.P. Levesque$^{1}$, G.A. Navratil$^{1}$, and M.E. Mauel$^{1}$\\[0.5em]
\small $^{1}$Department of Applied Physics and Applied Mathematics,\\
\small Columbia University, New York, NY, United States of America\\[0.3em]
\small \texttt{BL2718@columbia.edu}
}

\date{\today}
\maketitle

\begin{abstract}
This study examines the mechanisms underlying sawtooth suppression in the High Beta Tokamak-Extended Pulse (HBT-EP) device. It is observed that strong-intensity sawtooth activities correlate with reduced-amplitude MHD edge modes which are identified as $m/n=3/1$ external kink modes (XK), while sawtooth suppression correlates with larger and saturated edge mode amplitudes. To further investigate these correlations, the plasma-wall coupling was manipulated by adjusting the positions of the conducting walls in HBT-EP. It was found that strong sawtooth events occur when the normalized wall radius $b/a$ is within a critical value. This implies that the plasma-wall distance must be sufficiently small to ensure effective stabilization of the edge mode. Even slight differences in major radius result in significantly different discharge styles, categorized as ``sawtoothing discharges'' and ``sawtooth-suppressed discharges'' respectively. Through a series of mode structure analyses, we confirm the coexistence and coupling of the $m/n=1/1$ helical core (HC), $m/n=2/1$ tearing mode (TM), and $m/n=3/1$ XK during sawtooth suppression, and that this coupling induces anomalous current broadening. Based on these findings, we conclude that sawtooth suppression in the HBT-EP tokamak is consistent with the process of magnetic flux pumping. 
\end{abstract}

%
%
%
%
%

\section{Introduction}
The sawtooth instability is a recurring phenomenon in tokamak plasmas where the core plasma undergoes periodic redistributions, resulting in the cyclic reorganization of plasma density and temperature profiles \cite{Wesson97}. Sawteeth have been observed on various tokamaks since their initial discovery \cite{Goeler74} and have important implications for plasma performance. Frequent and small sawteeth can have beneficial effects, such as preventing impurity buildup in the plasma core and improving plasma confinement \cite{Nave03}. However, long-period sawteeth can trigger neoclassical tearing modes (NTMs) at low $\beta_N$ values, which can lead to disruptions and degradation of plasma confinement \cite{Sauter02, Gude02}. Therefore, controlling sawteeth is a crucial topic in fusion research. 

Sawtooth control refers to the ability of an actuator system to modify the period and amplitude of the sawtooth \cite{Graves05}, and can be achieved through various methods \cite{Chapman11}, including tailoring the distribution of energetic ions using techniques like ion cyclotron resonance heating (ICRH) or neutral beam injection (NBI) \cite{Berkery11, Berkery10}. It can also involve changing the radial profiles of plasma current density and pressure through techniques such as electron cyclotron current drive (ECCD) or electron cyclotron resonance heating (ECRH) \cite{Henderson01, Muck05, Maraschek05, Goodman11, Goodman11-2}.

Magnetic flux pumping involves controlling the formation of the $q=1$ surface through anomalous transport of plasma current, and is another method investigated for sawtooth control \cite{King12, Jardin15, Petty16, Petty17, Piovesan17, Na06}. It is one of the methods to achieve sawtooth-suppressed plasmas in hybrid scenarios\cite{Petty16, Petty17, Na06}. In DIII-D, flux pumping was observed with $n=1$ helical cores (HCs) induced by the $n=1$ magnetic perturbation\cite{Piovesan17}, or with NTM-ELM coupling\cite{Petty09}. JT-60U showed flux pumping through $n=1$ HC induced by $m/n=2/1$ NTM\cite{Bando19, Bando21PFR, Bando21PPCF}.

This paper focuses on the investigation of sawtooth suppression through flux pumping in the High Beta Tokamak-Extended Pulse (HBT-EP) device. We have observed that sawtooth suppression is associated with large and saturated edge mode amplitudes. In contrast, much lower amplitude edge modes are detected during strong sawtooth activity. Precise adjustment of the positions of the conducting walls in HBT-EP is used to control the presence of the edge mode. We show that the distance between the plasma and the wall must be within a critical value to effectively stabilize the edge mode, creating strong sawtooth events. Building upon this correlation, we further analyzed the mode components during sawtooth-suppressed periods and identified the coupling of the 1/1 HC, 2/1 internal tearing mode (TM) and the 3/1 external kink mode (XK). The coupling of these modes and the broadening of the current profile are the signatures of magnetic flux pumping. 

The article is organized as follows: Section 2 outlines the diagnostic and conducting wall systems in HBT-EP employed in this study. Section 3 reports our observations of sawtooth events and investigates how different wall locations lead to variations in discharge styles. Section 4 delves into the analysis of mode structures during sawtooth-suppressed periods to examine the role of the flux pumping mechanism. Section 5 uses an example discharge to illustrate the role of 1/1 HC in sawtooth suppression. Section 6 summarizes the study and presents the conclusions.

\section{Experimetal setup and diagnostics on HBT-EP}
HBT-EP is a tokamak device that features a high-aspect ratio configuration \cite{Maurer_2011}. It has a major radius of 92 cm and a minor radius of 15 cm. The tokamak is operated with a toroidal magnetic field $B_T$ of 0.33 T. The typical plasma current $I_p$ in HBT-EP ranges from 10 to 30 kA, and the pulse length of the discharges is usually between 6 to 10 ms. The discharge evolution and control in HBT-EP are achieved through pre-programmed capacitor banks. 

\begin{figure}
    {\includegraphics[width=.5\textwidth]{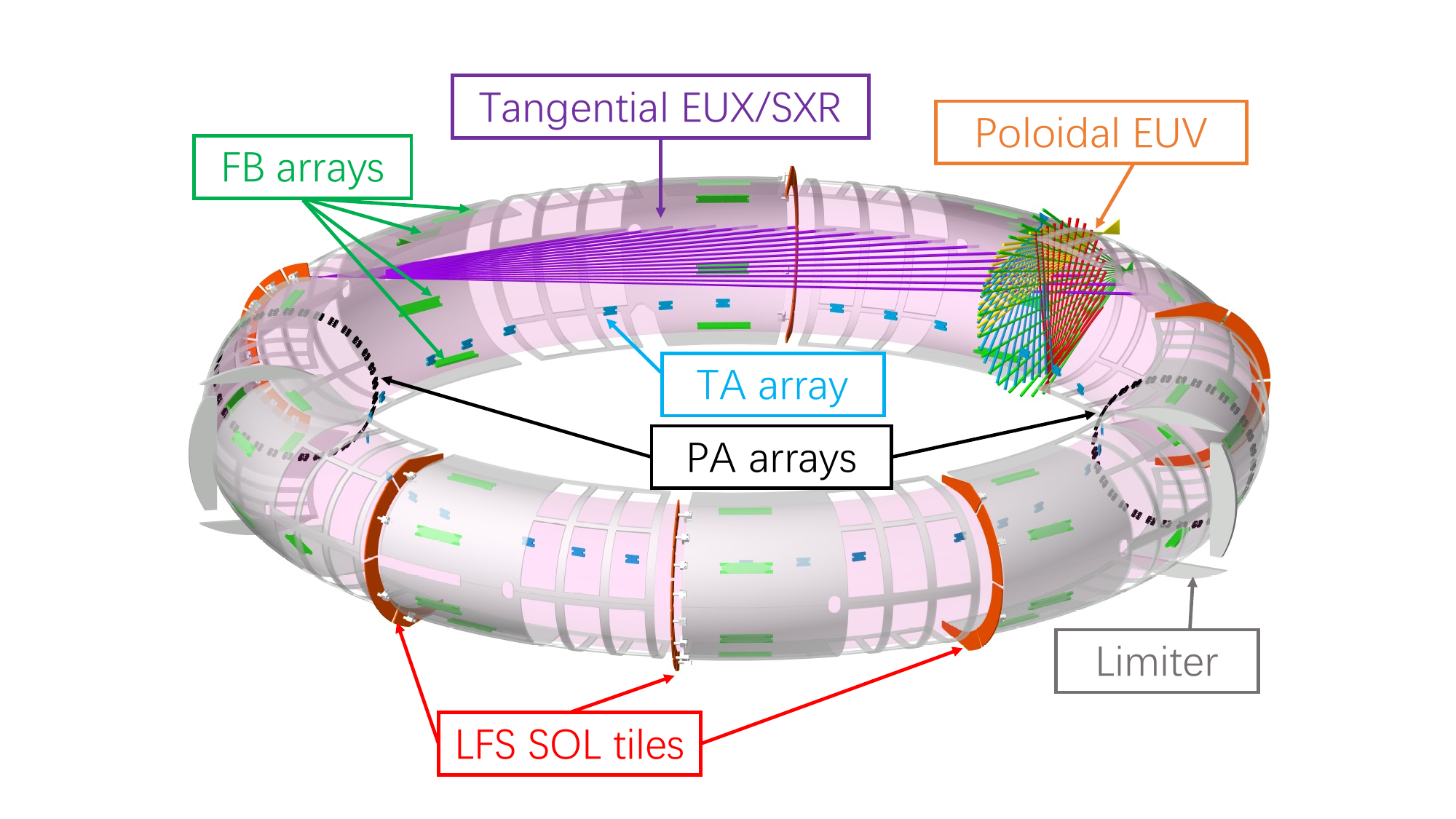}}
    \caption{Schematic of the HBT-EP tokamak, showing the tangential ME-EUV/SXR diagnostic system sightlines, poloidal EUV diagnostic system sightlines, magnetic diagnostic system, LFS SOL tiles, limiters and the conducting walls in their fully-inserted positions.}
    \label{fig:hbtep}
\end{figure}

The study employs a newly-installed tangential multi-energy extreme ultraviolet and soft x-ray (ME-EUV/SXR) diagnostic system on the HBT-EP device \cite{Li_2023}. Designed for low-temperature observations—below 200 eV—this diagnostic tool is well-suited for HBT-EP plasma conditions. The system features a filter wheel equipped with five distinct dual-filter sets and a two-stage amplifier with high gains reaching up to 60 M$\Omega$ and a bandwidth extending to 200 kHz. For measuring electron temperature, we utilize a filter combination of 0.1 $\mu$m aluminum and 0.2 $\mu$m titanium thin films. The line-integrated signals measured by two 16-channel photodiode arrays viewing the tangential mid-plane (marked in purple in Figure \ref{fig:hbtep}) are inverted to obtain plasma emission profiles, which in turn allows for the measurement of electron temperature. The signals are smoothed using a first-order low-pass Butterworth filter to reduce high-frequency noise. The cutoff frequency is set at 12 kHz, with the forward-backward (filtfilt) method applied to ensure a zero-phase-shift response. 

The poloidal EUV diagnostic system has four photodiode arrays, labeled 1 to 4 in a counterclockwise manner \cite{Chandra_2023}. As shown in Figure \ref{fig:hbtep}, each detector array consists of 16 diode channels that are strategically positioned to provide various sightlines across the poloidal cross-section of the plasma. To ensure the detection of EUV radiation specifically, the system employs 0.1 $\mu$m aluminum thin films as filters. In this study, we use the poloidalEUV array at the mid-plane, with a vertically-symmetric view of the plasma.  Chords are labeled 1 through 16, starting from the view of the bottom of the plasma.

The magnetic diagnostic system employed in the HBT-EP device is composed of a total of 216 sensors constructed from Mirnov coils. These sensors are organized into three distinct sets of arrays and enable high-resolution and high-sensitivity measurements of magnetic fluctuations that arise from edge modes. The first set is the toroidal array (TA), which is an array of Mirnov coils distributed on the high-field side of the plasma. The second set, the feedback (FB) arrays, consists of sensors arranged in a 4$\times$10 configuration and provides measurement of edge mode dynamics. Finally, the third set comprises the two poloidal arrays (PA), which are located 180 degrees apart toroidally and are distributed evenly around the poloidal cross section. The magnetic array data can be used in singular value decomposition (SVD) analysis to study the mode structures in the plasma.  

\begin{figure}
    {\includegraphics[width=.45\textwidth]{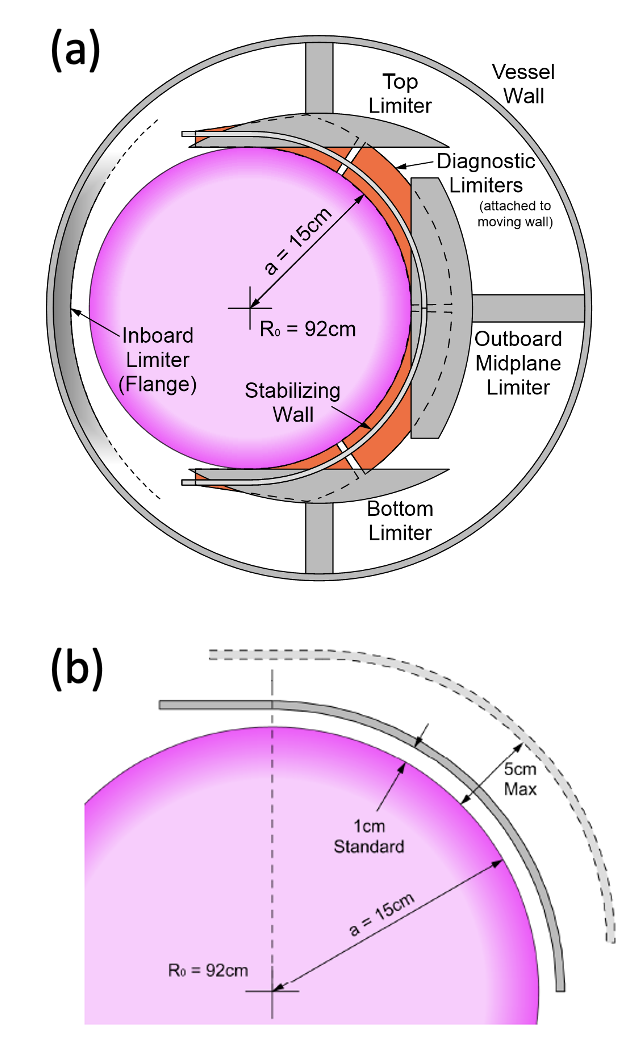}}
    \caption{(a) The schematic of the LFS SOL tile, chamber wall, limiters and the conducting wall. (b) The moving range of the conducting wall. When each shell is fully inserted, its inner face is positioned 1 cm away from the plasma surface, assuming a plasma centered at $R_0=$ 92 cm with $a=$ 15 cm. Individual shells can be retracted up to 5 cm from the plasma surface along a 45$^{\circ}$ angle, as shown by the dashed shell outline.}
    \label{fig:shells}
\end{figure}

In HBT-EP experiments, the plasma edge is defined by the low-field-side scrape-off layer (LFS SOL) tiles and the limiters as shown in Figure \ref{fig:shells}a. The LFS SOL tiles have a minor radius of 15 cm and a major radius of 92 cm when positioned at their innermost location. The top and bottom limiters are located at $Z$ = $\pm$15 cm, the outboard limiter sits at $R$ = 107 cm, and the inboard limiter sits at $R$ = 75 cm. When the plasma major radius $R_0$ is between 90 to 92 cm, the plasma is confined to have a minor radius $a$ = 15 cm. When $R_0$ is outside this range, $a$ is defined as $R_0$ - 75 cm ($R_0<$ 90 cm) or 107 cm - $R_0$ ($R_0>$ 92 cm).

In this study, we move the 316 SS conducting wall segments to manipulate the behavior of the edge mode \cite{Maurer_2011}. The system, located on the low-field side, consists of 20 independent wall segments. Each segment is divided into three window-pane cutouts arranged in two rows, along with a continuous portion. The continuous portion is coated with a copper layer through electroplating and serves as a passive stabilizer for ideal external kink and resistive wall modes \cite{Garofalo_1998, Maurer_2011, Ivers_1996}. The copper layer has a nominal thickness of 5 mil (2.5 mil on each side of the wall), selected based on VALEN modeling \cite{Bialek_2001}, to achieve a wall time of approximately 400 microseconds for the $n=1$ RWM. To allow for flexibility in plasma-wall coupling, each wall can be independently radially adjusted over a range of 4 cm using an actuator. When fully inserted, the inner face of each wall resides 1 cm from the plasma surface, assuming a plasma centered at $R_0=$ 92 cm with $a=$ 15 cm (Figure \ref{fig:shells}b). This unique movable wall configuration enables precise modulation of the interaction between the plasma and the conducting walls.

A key parameter manipulated in this study is the plasma-wall distance. To mark the location of the conducting walls, $D$ is defined to represent the distance between the inner surface of the wall and the plasma surface when the plasma is centered at major radius $R_0 =$ 92 cm. The absolute distance between the wall and the center of the device can be calculated as $s = 107$ cm + $D$. To describe the plasma-wall dynamic, we introduce the parameter $d$, which depicts the distance between the wall's inner surface and the plasma surface, formulated as $d = s-(R_0 + a)$. The conducting wall's radius is defined as $b = a + d$. The normalized wall radius $b/a$ is a crucial metric to represent the wall location and study the plasma-wall interaction \cite{Wesson_1978, Cates_2000}.

\section{Bifurcation of sawtooth behavior by wall stabilization of edge mode}
\subsection{Observation of sawtooth}
\begin{figure}
    \centering
    \subfloat[]{\includegraphics[width=.45\textwidth]{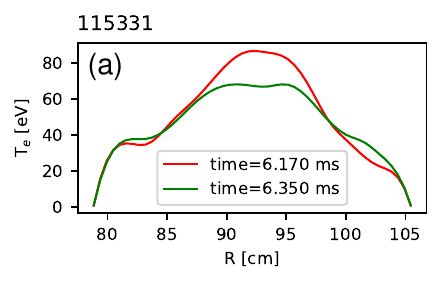}\label{fig:crosssection}}
    \subfloat[]{\includegraphics[width=.45\textwidth]{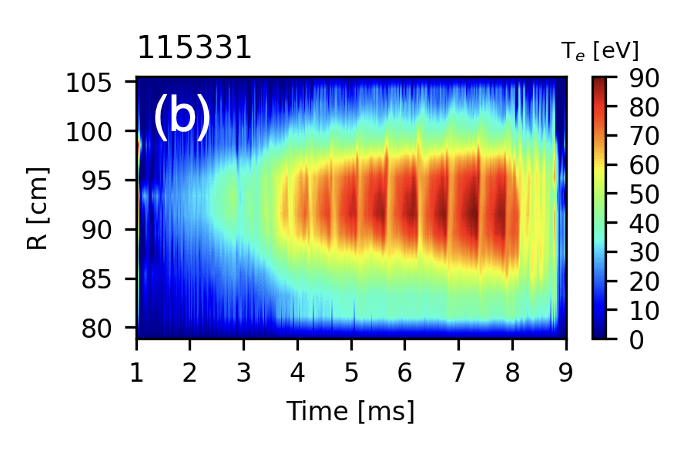}\label{fig:sawtooth}}
    \\
    \subfloat[]{\includegraphics[width=.8\textwidth]{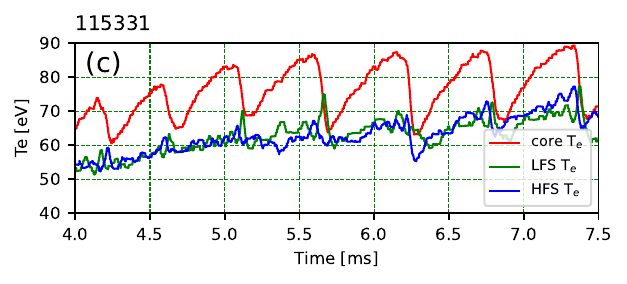}\label{fig:single_115331}}
    \caption{Sawtoothing discharge 115331: (a) Electron temperature profile measured before (red) and after (green) a sawtooth crash. (b) The time evolution of the electron temperature profile. (c) Temperatures in the core and mid-radius areas on the low-field side (LFS) and high-field side (HFS). $R=96.8$ cm for LFS T$_e$, $R=88.1$ cm for HFS T$_e$ and $R=91.9$ cm for core T$_e$.}
    \label{fig:115331}
\end{figure}

The tangential EUV/SXR system has allowed for the first detailed observations of sawtooth events in the core of HBT-EP tokamak plasmas. Using the signals acquired from the 16 channels of each diode array at each time-point, we can reconstruct the radial electron temperature profile \cite{Li_2023}. Figure \ref{fig:crosssection} illustrates the resulting radial profile at two time points in an example discharge 115331. During a rapid event, the electron temperature profile undergoes significant changes, transitioning from a peaked to a flat shape while also expanding to a wider minor radius. These observations indicate the occurrence of a temperature crash and the existence of an inversion radius, both of which are characteristic features of sawtooth events.

The electron temperature profile over time can be reconstructed using the signals during a given discharge. The results for discharge 115331 is depicted in Figure \ref{fig:sawtooth}. The core and mid-radius electron temperatures are shown in Figure \ref{fig:single_115331}. The periodic and repetitive process of peaking and collapsing is shown. These observations provide strong evidence for the occurrence of sawtooth events.

\subsection{Wall stabilization of edge mode and its impact on sawtooth suppression}
In our study, we also came across certain discharges where sawtooth activity was suppressed. Despite minor changes in plasma current and major radius, these discharges do not display strong periodic peaking and collapsing behaviors.

\begin{figure}
    \centering
    \includegraphics[width=.95\textwidth]{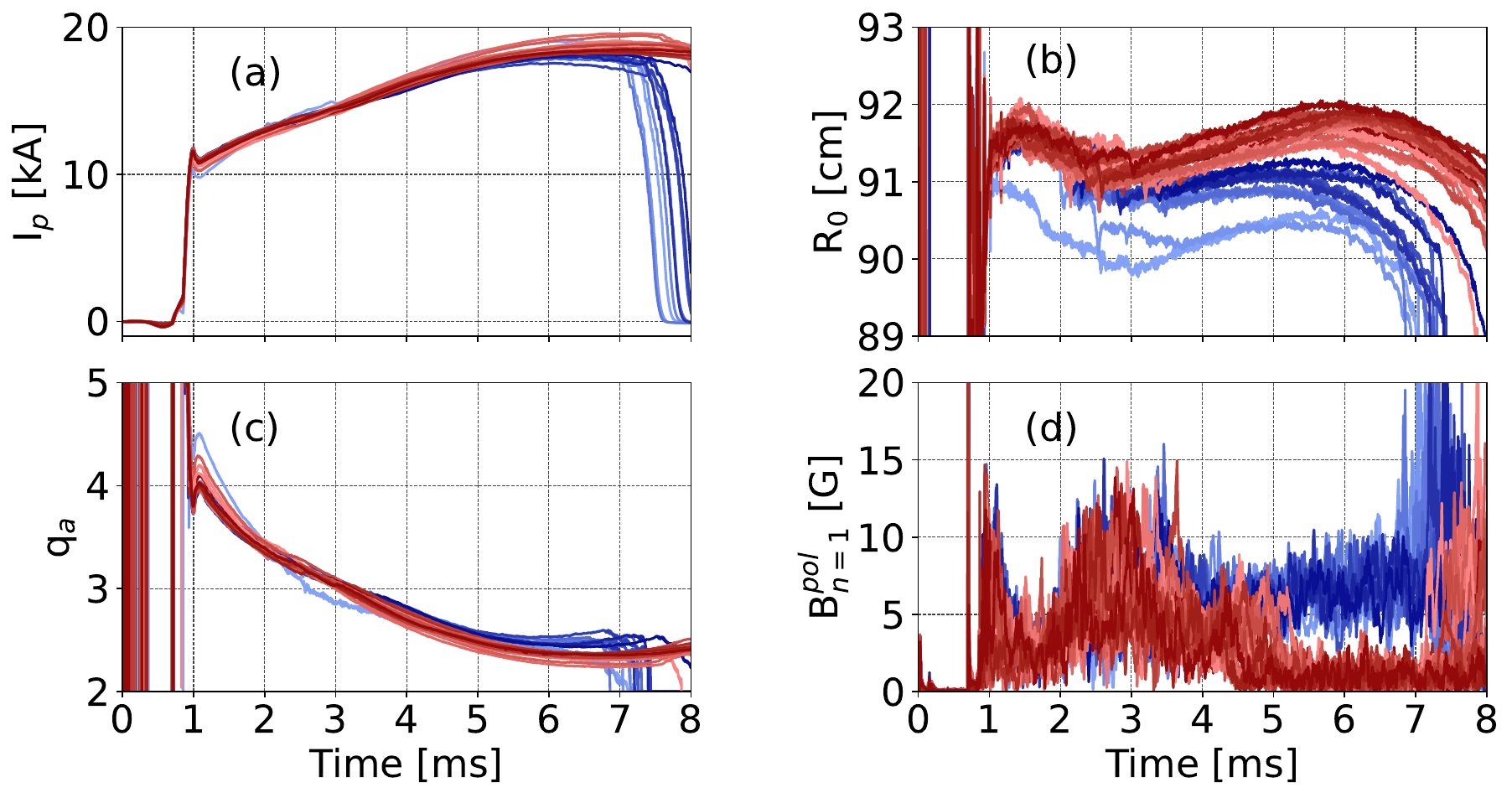}
    \caption{Comparison of plasma parameters between 18 sawtoothing discharges (red) and 11 sawtooth-suppressed discharges (blue) when $D=$ 10 mm. The consistent and repetitive characteristics displayed by these discharges imply the repeatability of this phenomenon. (a) The plasma currents. (b) The major radii. (c) The edge safety factors. (d) The edge mode amplitudes.}
    \label{fig:summary}
\end{figure}

Figure \ref{fig:summary} summarizes the plasma parameters of 18 sawtoothing discharges (red curves) and 11 sawtooth-suppressed discharges (blue curves) when all conducting walls are fully inserted ($D$ = 10 mm). It can be seen that sawtooth activities are associated with outboard major radii and low-amplitude edge modes. In contrast, they are suppressed when major radii are inboard and the edge modes exhibit large and saturated amplitudes. The consistent characteristics of these discharges suggest the phenomenon is repeatable. These differences in major radii and mode amplitudes between the two discharge styles indicate that the interaction between the plasma and the conducting walls play a significant role in determining their divergent sawtooth behaviors. 

Systematic experiments were conducted altering the plasma-wall distance by adjusting the wall positions. We varied the wall positions to $D$ = 10 mm, 12.5 mm, 15 mm, 17.5 mm, 20 mm, and 25 mm while maintaining similar plasma parameters. The experiment results establish a direct and clear relation between the amplitude of edge modes, the occurrence of sawtooth events, and a critical major radius $R_c$ in HBT-EP plasmas (Table \ref{table_dc}). For each wall position with $D\le15$ mm, when the major radius exceeds $R_c$, strong sawtooth events occur. Conversely, when the major radius is below $R_c$, sawtooth events are suppressed. This critical radius varies based on the positions of the walls. Increasing $D$ results in an elevation of $R_c$. 

\begin{figure}
    \centering
    \subfloat[]{\includegraphics[width=.95\textwidth]{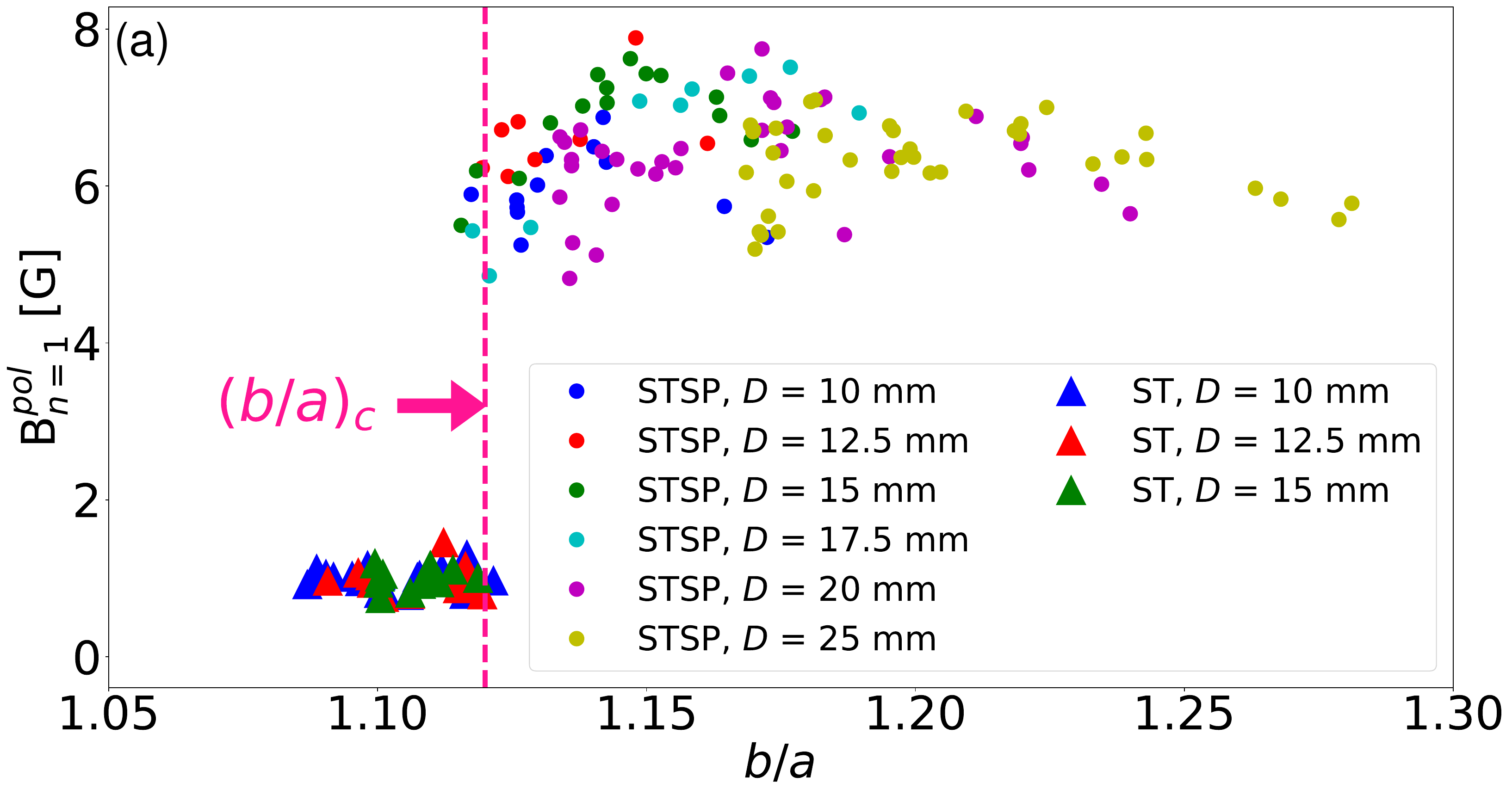}\label{fig:wall_scan_ba}}
    \\
    \subfloat[]{\includegraphics[width=.95\textwidth]{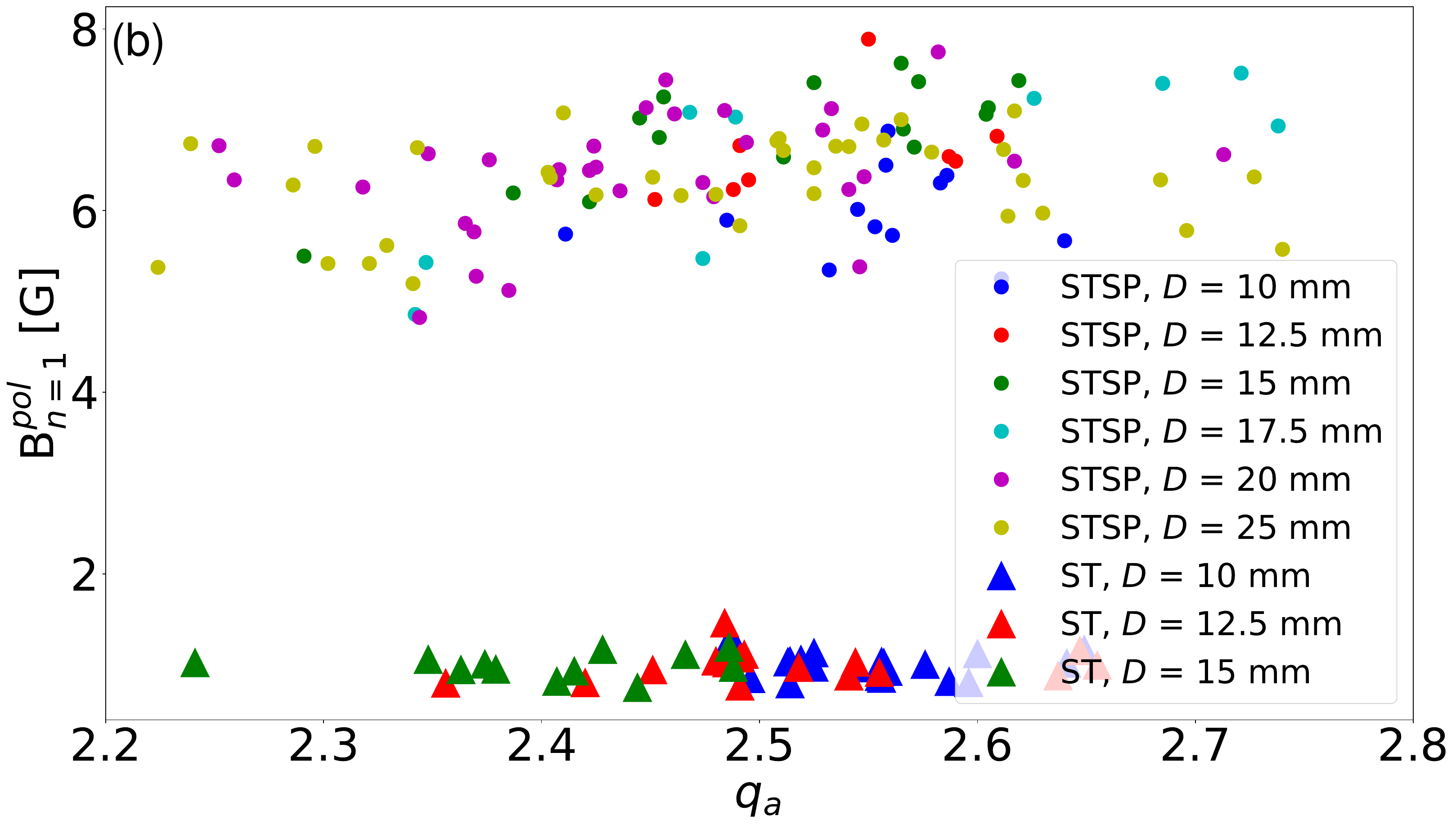}\label{fig:wall_scan_q}}
    \caption{For each wall location: (a) the edge mode amplitude versus $b/a$ and (b) the edge mode amplitude versus $q_a$. $(b/a)_c$ is marked by the dashed line. Sawtooth-suppressed discharges (STSP) are represented by circles, and sawtoothing discharges (ST) are represented by triangles. For sawtoothing discharges, $b/a$ values are calculated at the time point when the edge mode begins to decay. For sawtooth-suppressed discharges, $b/a$ values are calculated at the time point with maximum major radius, which correspond to minimum $b/a$. }
    \label{fig:bifurcation}
\end{figure}

Based on the observed results, the critical plasma-wall distance $d_c$ for sawtooth occurrence is 17.8$\sim$18.5 mm (approximately 12$\%$ of the minor radius). Correspondingly, the critical value of the normalized wall radius $(b/a)_c$ is found to be 1.12. As shown in Figure \ref{fig:wall_scan_ba}, this critical $b/a$ value is independent of the wall locations given the current measurement precision. Maintaining $b/a<(b/a)_c$ is crucial for effective stabilization of the edge mode, allowing the occurrence of strong sawtooth instability. Figure \ref{fig:wall_scan_q} indicates that the stabilization of the edge mode during the period of interest in these HBT-EP plasmas is not correlated to variations in the edge safety factor. This observation rules out the possibility that changes in the $q$ profile are responsible for the stabilization of the edge mode. These findings highlight the significant role played by the distance between the plasma and the surrounding walls in influencing the dynamics of the edge mode and the sawtooth instability in the HBT-EP experiment.

Observations indicate that for effective stabilization of the edge modes, and hence the occurrence of strong sawtooth events, $D$ must be within 15 mm. By comparison, when $D$ = 17.5 mm or larger, the edge mode saturates with a large amplitude and sawtooth suppression occurs, as shown in Figure \ref{fig:wall_scan_ba}. This particular behavior can be attributed to the presence of the outboard limiter and LFS SOL tiles in HBT-EP, which restrict plasma expansion and prevent the plasma-wall distance from decreasing. $b/a$ maintains above the critical value $(b/a)_c$ needed for effective edge mode stabilization. As a result, if the walls are retracted too far, they cannot stabilize the edge mode, even with an outboard plasma major radius.

\begin{table}
\centering

\lineup
\begin{tabular}{@{}*{4}{c}}
\br                              
$D$&$R_c$&$d_c=s-(R_c+a)$&$(b/a)_c=(a+d_c)/a$\\ 
\mr
10 mm & 912.2 mm & 17.8 mm & 1.12\\
12.5 mm & 914.4 mm & 18.1 mm & 1.12\\
15 mm &  916.5 mm & 18.5 mm & 1.12\\
\br
\end{tabular}
\caption{The critical value of the normalized wall radius for sawtooth occurrence $(b/a)_c$ is independent of wall locations within the current level of precision.} 
\label{table_dc}
\end{table}

\begin{figure}
    \centering
    \subfloat[]{\includegraphics[width=.4\textwidth]{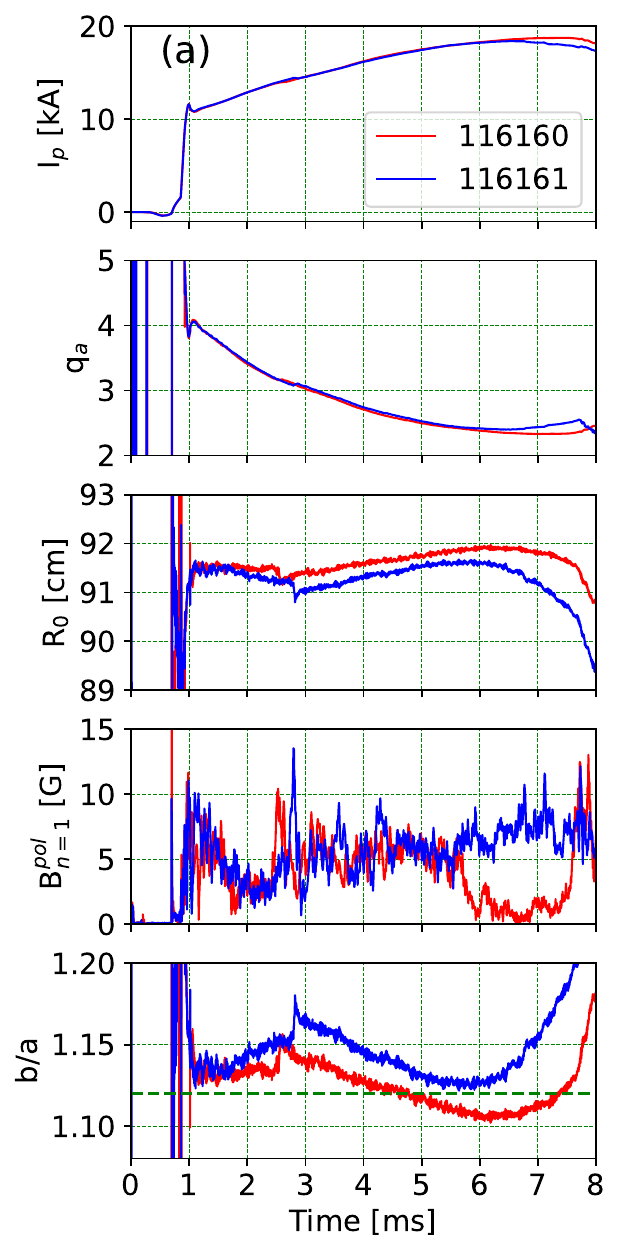}\label{fig:parameters}}
    \subfloat[]{\includegraphics[width=.5\textwidth]{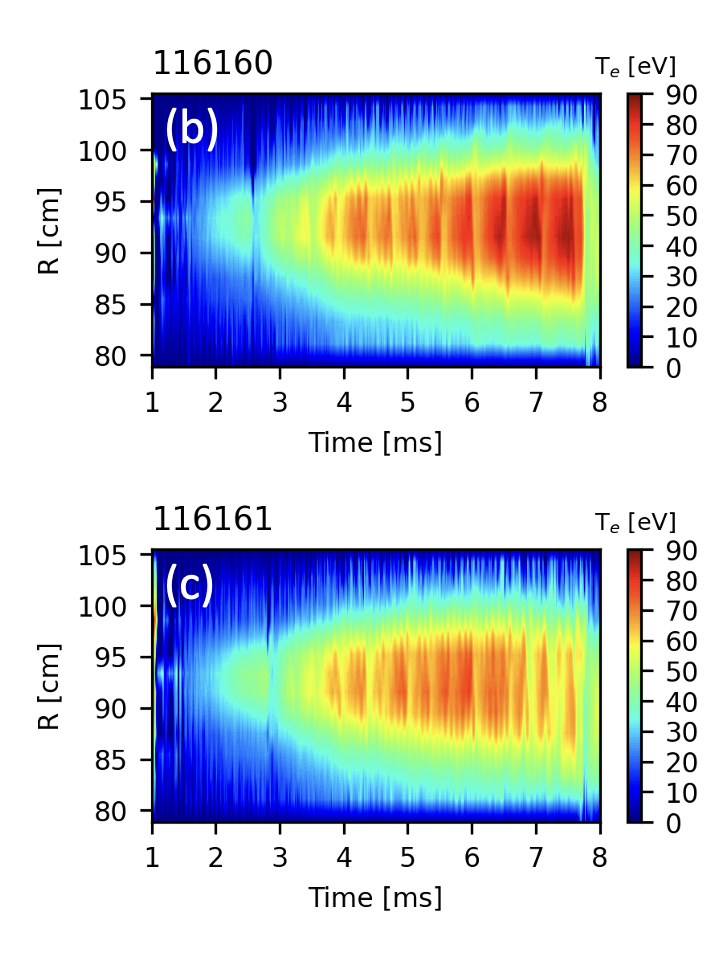}\label{fig:Te_compare}}
    \caption{Slight change in plasma parameters leads to discharge style bifurcation in successive discharges. The critical normalized wall radius $(b/a)_c=1.12$ is marked by the green dashed line in the bottom left plot.}
    \label{fig:116160&116161}
\end{figure}

Based on HBT-EP geometry, $b/a$ can be modified by either moving the walls or plasma major radius. Therefore, the impact of plasma-wall interactions on sawtooth events can be observed when the major radius varies over time without altering the physical wall position. This is illustrated by comparing discharges 116160 and 116161 in Figure \ref{fig:116160&116161}. Both discharges have comparable plasma current and edge safety factors, but their major radii differ marginally. Notably, their electron temperature profiles exhibit distinct characteristics. As shown in Figure \ref{fig:parameters}, in discharge 116160, the wall stabilizes the edge mode when the major radius $R_0$ moves outboard, reducing $b/a$ below $(b/a)_c$. This results in stronger sawtooth activities (Figure \ref{fig:116160&116161}b). In contrast, discharge 116161 is classified as a ``sawtooth-suppressed'' discharge due to its stable state without strong sawtooth events (Figure \ref{fig:116160&116161}c). This discharge maintains a consistent, saturated edge mode amplitude throughout. With its more inboard $R_0$, $b/a$ remains above $(b/a)_c$, preventing effective edge mode stabilization by the wall, hence suppressing sawtooth. Additional individual discharges with more extreme $b/a$ values are shown in the Appendix.

These observations demonstrate that the plasma-wall distance has a notable effect on the amplitude of the edge mode through plasma-wall interactions. These interactions play a critical role in stabilizing the edge mode and determining the occurrence of sawtooth events. In the following section, we will provide a comprehensive analysis of the mode structures during the sawtooth-suppressed stage, aiming to gain a deeper understanding of the underlying mechanism driving the suppression of sawtooth.

\section{Mode analysis of sawtooth-suppressed phenomena}
To uncover the underlying mechanism responsible for the suppression of sawtooth phenomena, we focus on a more in-depth exploration of the modes present in the sawtooth-suppressed period and their interactions. To find evidence of the existence of various modes, we examine the plasma behavior through the edge to the core. 

\begin{figure}
    \centering
    \subfloat[]{\includegraphics[width=.7\textwidth]{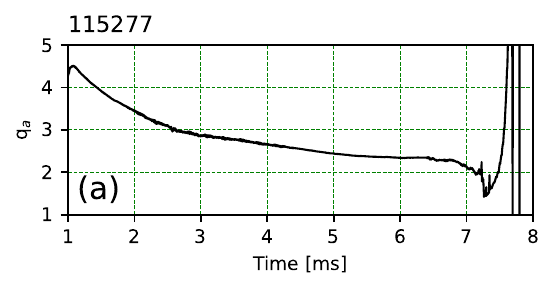}\label{fig:q_115277}}
    \\
    \subfloat[]{\includegraphics[width=.6\textwidth]{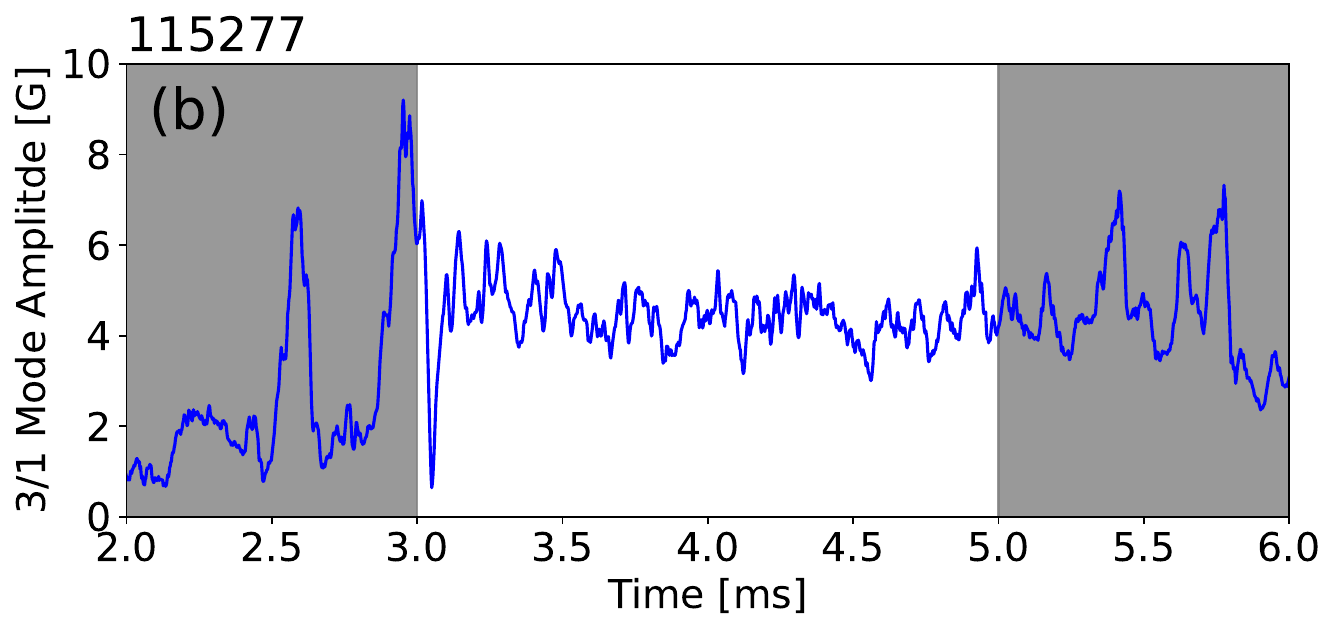}\label{fig:115277_modeamp}}
    \subfloat[]{\includegraphics[width=.3\textwidth]{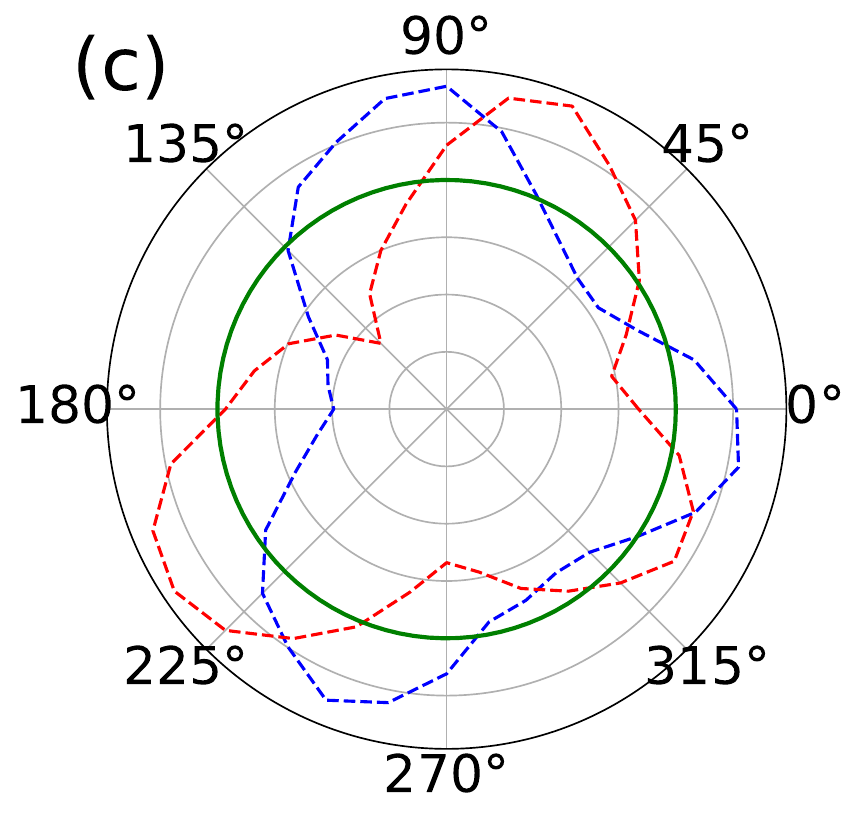}\label{fig:SVD_115277_3}}

    \caption{Evidence of 3/1 XK in discharge 115277: (a) The edge q is below 3 after 3 ms. (b) Mode amplitude of the 3/1 mode. The 3/1 mode remains saturated during 3-5 ms. (c) The SVD analysis result of PA sensor data during 3-5 ms shows 3/1 components. The blue and red curves are the sine and cosine SVD spatial modes corresponding to the 3/1 mode. The radial location marked in green represents the reference center of fluctuation. }
    \label{fig:115277_3}
\end{figure}

\begin{figure}
    \centering
    \subfloat[]{\includegraphics[width=.7\textwidth]{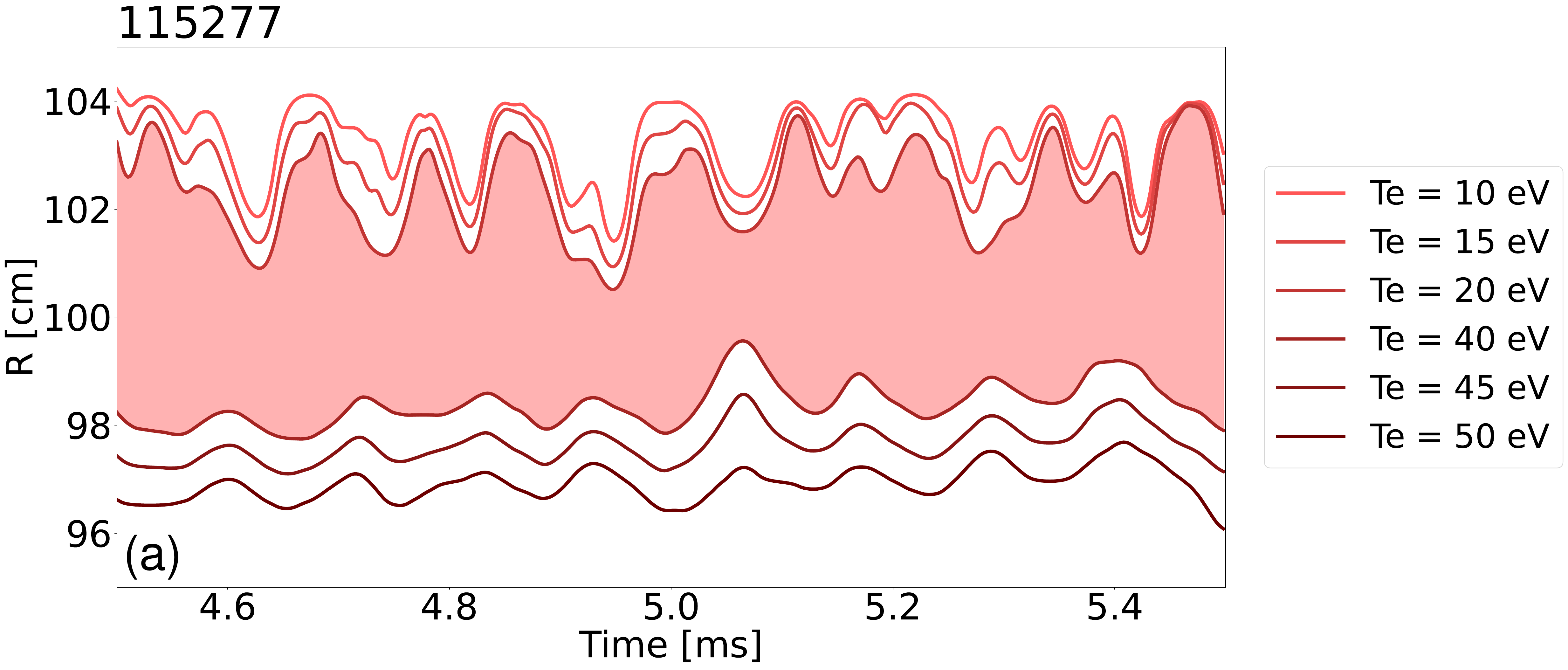}\label{fig:Tearing_115277}}
    \\
    \subfloat[]{\includegraphics[width=.8\textwidth]{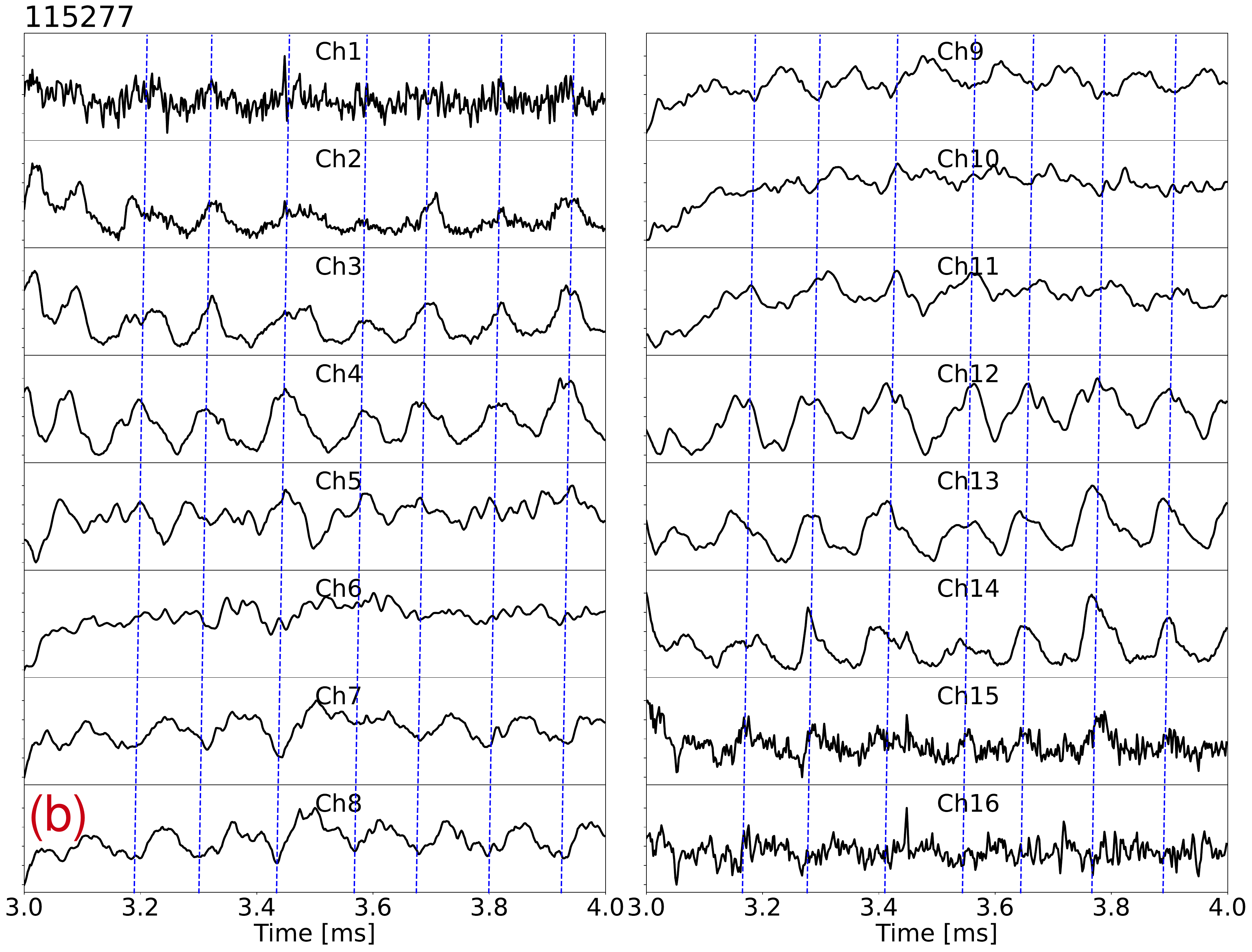}\label{fig:pol_115277}}
    \caption{Evidence of 2/1 TM in discharge 115277: (a) The $T_e$ contours show inversion of the temperature fluctuations. (b) Signals from the mid-plane array of the poloidal EUV system show phase inversions at Ch 6 and Ch 10. }
    \label{fig:115277_2}
\end{figure}

\begin{figure}
    \centering
    \subfloat[]{\includegraphics[width=.8\textwidth]{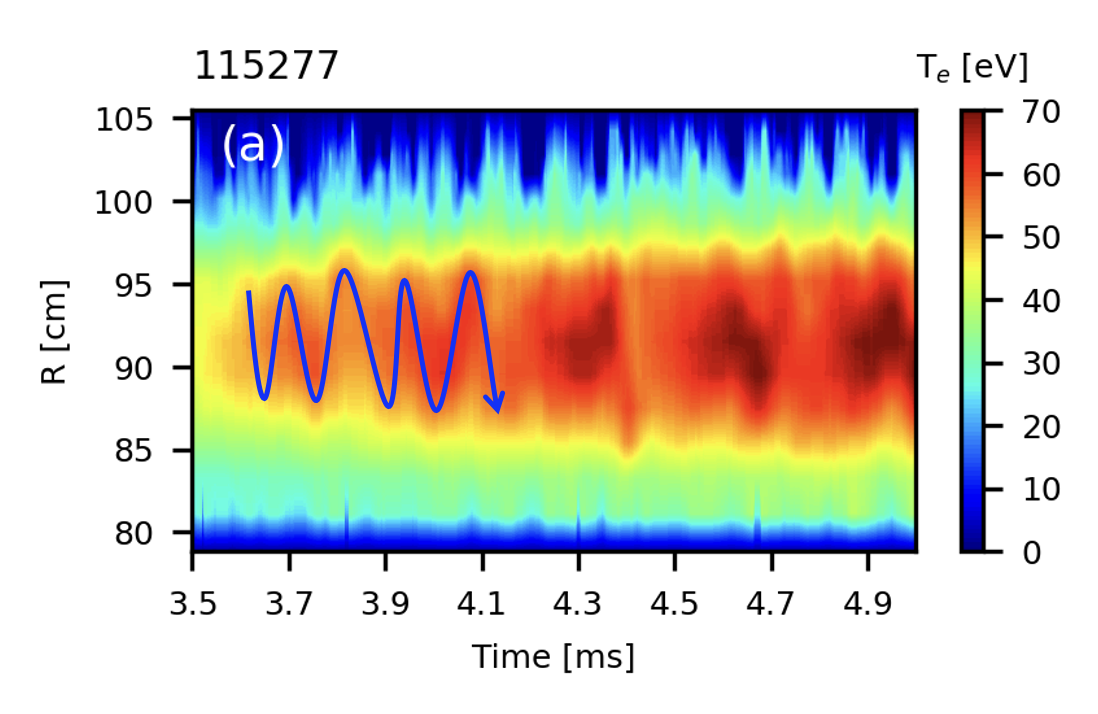}\label{fig:snake_115277}}
    \\
    \subfloat[]{\includegraphics[width=.7\textwidth]{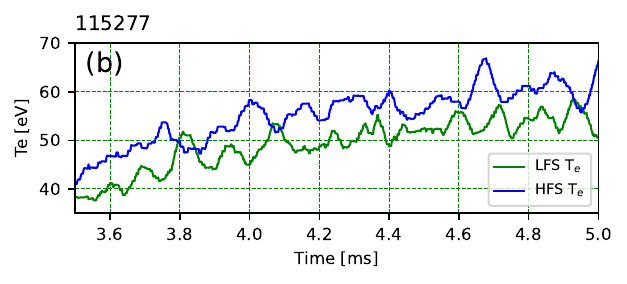}\label{fig:single_115277}}
    \caption{Evidence of 1/1 HC in discharge 115277: (a) Rotating snake structure in the core. (b) The radial temperature measurements show an out-of-phase feature. $R=96.2$ cm for LFS T$_e$ and $R=88.1$ cm for HFS T$_e$.}
    \label{fig:115277_1}
\end{figure}

To study the mode structures and their interactions, we choose discharge 115277 as a representative example of sawtooth-suppressed discharges. Based on results presented in the previous section, it's evident that the conducting walls can stabilize the edge mode. Given that the edge safety factor $q$ is below 3 after 2.5 ms and the mode has an $m/n=3/1$ structure on the external magnetic sensors as shown in Figure
\ref{fig:q_115277}, this mode is categorized as an XK. The presence of the 3/1 XK is corroborated by the SVD analysis of the PA array data which reveals a 3/1 structure during 3-5 ms (Figure \ref{fig:SVD_115277_3}). This kink mode saturates and persists without immediately disrupting the plasma, as is often seen in HBT-EP plasmas \cite{Levesque_2013,Levesque_2015}. As shown in Figure \ref{fig:Tearing_115277}, the observed phase inversions in the temperature contour plots indicate the existence of a tearing mode. Moreover, we can see from Figure \ref{fig:pol_115277} that the mid-plane array signals from the poloidal EUV system present a distinct inversion pattern, characterized by two inversions across the poloidal cross section of the plasma. The first is observed at Ch6, which has a impact parameter (the perpendicular distance from the plasma center to the sightline of the corresponding channel) of -58.4 mm. The second is observed at Ch 10, with an impact parameter of +43.9 mm. This pattern further indicates the presence of a 2/1 TM. Furthermore, insights from both the snake-like pattern found in the temperature measurements (Figure \ref{fig:snake_115277}) and the out-of-phase pattern in the radial temperature results (Figure \ref{fig:single_115277}) suggest the presence of a 1/1 helical core (HC).

\begin{figure}
\centering
\includegraphics[width=0.8\textwidth]{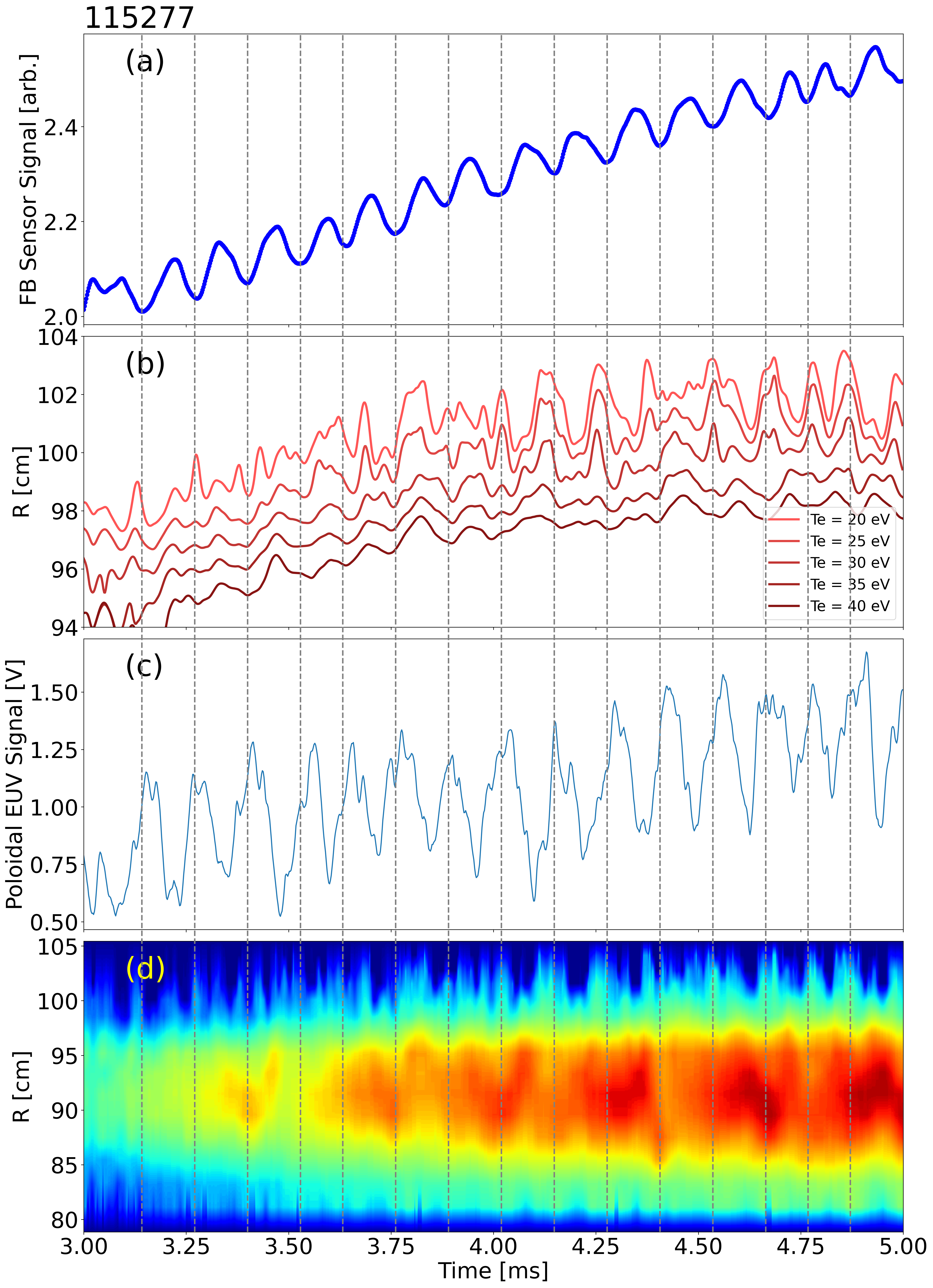}
\caption{\label{fig:115277_cp} Mode coupling in discharge 115277: (a) FB sensor signal representing 3/1 XK amplitude. (b) Electron temperature contours representing 2/1 TM. (c) Signal of poloidal EUV system mid-plane array channel 12 representing 2/1 TM. (d) Snake structure in the core electron temperature measurement representing 1/1 HC. The phase matching among these results indicate mode coupling among 3/1 XK, 2/1 TM and 1/1 HC.}
\end{figure}

Importantly, correlation between the external magnetic signals and the internal EUV/SXR signals has shown the coupling among the modes. The 3/1 XK dynamics can be represented by the FB sensor signals since they provide information on edge mode activities. Both the FB sensor readings (Figure \ref{fig:115277_cp}a) and the temperature contour plot (Figure \ref{fig:115277_cp}b) exhibit phase coherence, suggesting mode locking between the 3/1 XK and the 2/1 TM. The signal from the poloidal EUV system (Figure \ref{fig:115277_cp}c) aligns in phase with that of the FB sensor, which further supports the mode locking between the 3/1 XK and the 2/1 TM. The snake-like pattern identified in the temperature measurements (Figure \ref{fig:115277_cp}d) aligns in phase with the FB sensor data, which reveals phase locking between the 3/1 XK and the 1/1 HC. This mode locking mechanism is potentially important in the observed suppression of sawtooth activity in these discharges.

The conducted analyses reveal the concurrent existence and coupling of the 3/1 XK, 2/1 TM, and 1/1 HC during the period of sawtooth suppression in HBT-EP plasma, and this suggests a similar mechanism to the phenomenon of magnetic flux pumping reported in DIII-D \cite{Petty09}. For DIII-D, the coupling of an internal mode (NTM) with an external mode (ELM) results in the mitigation of sawtooth instability; whereas for HBT-EP, the coupling involves an XK.

Central to the mechanism of sawtooth suppression via magnetic flux pumping is the 1/1 HC. This component is instrumental in initiating the dynamo loop voltage, which in turn leads to anomalous transport of magnetic flux and broadening of the current profile, thereby sustaining central safety factor $q_0$ around unity \cite{Krebs17, Jardin15, Piovesan17}. The origins of the 1/1 HC differ across devices. In the context of HBT-EP, it is likely that the 1/1 components arising from the 3/1 XK and 2/1 TM collectively give rise to the 1/1 HC. 

\begin{figure}
\centering
\includegraphics[width=0.9\textwidth]{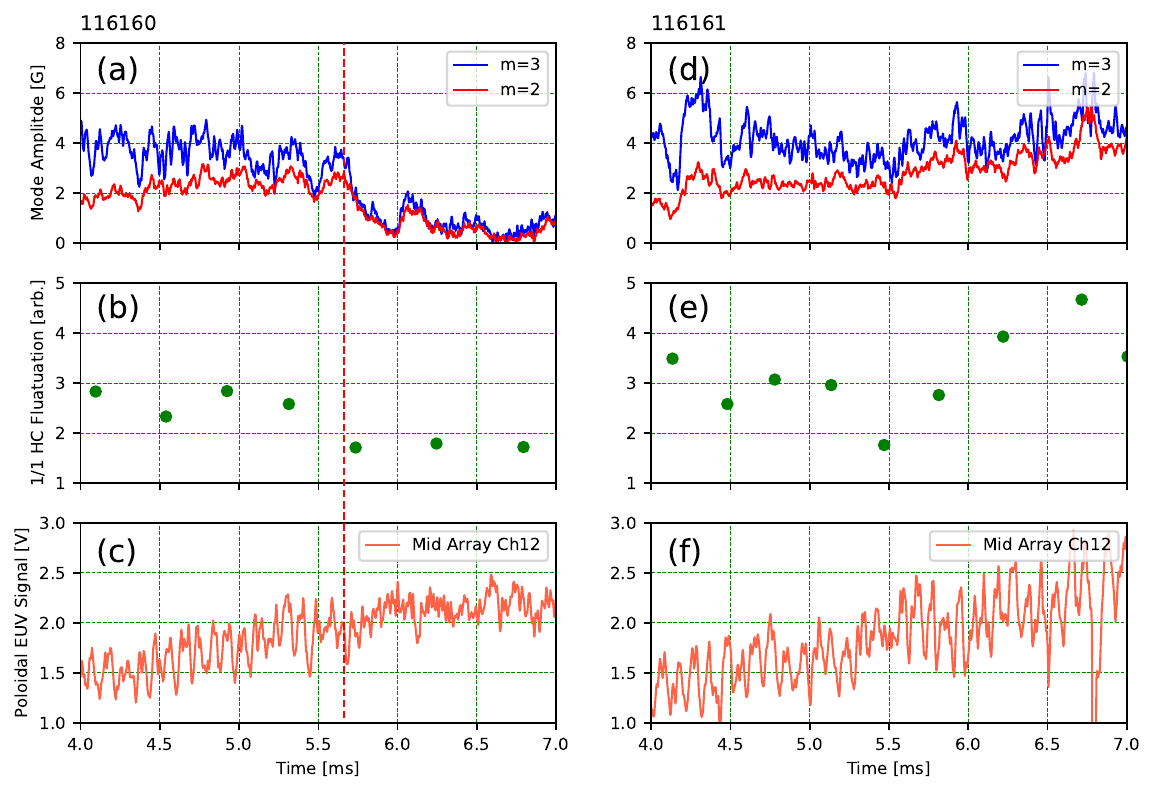}
\caption{\label{fig:Fourier_mode} (a) and (d): Amplitudes of the 3/1 XK and 2/1 TM calculated through spectral analysis using PA signal fluctuations. (b) and (e): Amplitude of the 1/1 HC determined by analyzing core fluctuations in titanium signals from the tangential EUV/SXR system. (c) and (f): Amplitudes of the 2/1 TM illustrated by fluctuations in signal from channel 12 of the poloidal EUV mid array. In discharge 116160, as the 3/1 XK is stabilized by the conducting wall, a corresponding stabilization in the 1/1 HC is observed, along with a decrease in the 2/1 TM amplitude. In contrast, in discharge 116161, the 3/1 XK remains at a saturated level, leading to stable amplitudes of both the 1/1 HC and the 2/1 TM.}
\end{figure}

In our study, we analyzed titanium signals from the tangential EUV/SXR system to quantify the 1/1 HC-induced fluctuations. We performed Abel inversion on these signals and selected four central channels to represent core plasma dynamics. During the rise period of the sawtooth, we applied a linear fit to each channel, then subtracted the fit to extract fluctuations. The fluctuation amplitudes were determined using relative standard deviation, calculated as the standard deviation divided by the mean for each channel. These amplitudes were then averaged across the chosen channels to quantify the fluctuations for each sawtooth period. In Figure \ref{fig:Fourier_mode}, the spectral analysis results of 3/1 and 2/1 components' amplitudes from PA fluctuations (including compensation for toroidal geometry per the method described in \cite{Kluber_1991}), 1/1 HC fluctuations quantified using the titanium signals from tangential EUV/SXR system, and 2/1 TM fluctuations represented by the signals from poloidal EUV system are compared. When the 3/1 XK becomes stabilized by the conducting wall, we observed simultaneous stabilization of the 2/1 TM and the 1/1 HC. This observation indicates that the wall stabilization of the 3/1 XK leads to an overall reduction of all three modes, and the 2/1 TM is likely to be induced by the 3/1 XK.

\section{Role of 1/1 HC in sawtooth suppression}
As we discussed in the previous section, 1/1 HC is the key component that induces dynamo loop voltage and increases $q_0$ during the sawtooth suppression stage. By examining the electron temperature measurements at the core and $q=1$ surface, we can clearly see how the existence of 1/1 HC affects the sawtooth occurrence. 

\begin{figure}
\includegraphics[width=0.9\textwidth]{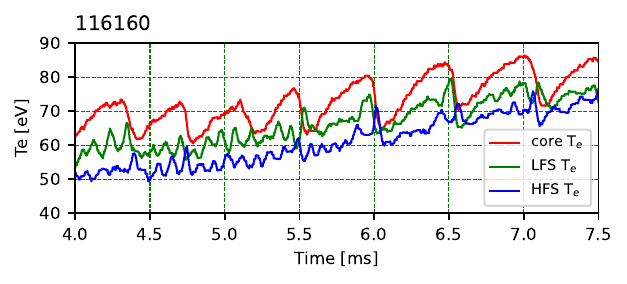}
\caption{\label{fig:116160_single} Discharge 116160: Electron temperature of the core and mid-radius area on the LFS and HFS. $R=96.2$ cm for LFS T$_e$, $R=88.1$ cm for HFS T$_e$ and $R=92.4$ cm for core T$_e$. The 3/1 XK decays at 5.7 ms, yielding stronger sawteeth.}
\end{figure}

Figure \ref{fig:116160_single} presents discharge 116160 as an illustrative case. As shown in Figure \ref{fig:116160&116161}, the changing major radius in this discharge results in variation in the 3/1 XK amplitude and corresponding sawtooth behavior. Notably, prior to 5.5 ms as shown in Figure \ref{fig:116160&116161}b, current broadening, which is a characteristic of magnetic flux pumping, is evident during the sawtooth-suppressed phase, as highlighted by the noticeable expansion of the hot region.

As shown in Figure \ref{fig:116160_single}, before the decrease in edge mode amplitude around 5.5 ms, the core temperature measurement indicates weaker sawtooth activity, which can be attributed to the effects of flux pumping. Simultaneously, the electron temperature measurements at the mid-radius manifest an out-of-phase pattern when comparing results from low-field side and high-field side, revealing the presence of a snakelike 1/1 HC . 

After 5.7 ms, the 3/1 XK amplitude starts to decrease, influenced by the stabilizing effect of the conducting walls as the major radius moves outboard. The decline in the 1/1 mode amplitude is clearly depicted in Figure \ref{fig:116160_single}, illustrated by the disappearance of the out-of-phase pattern. Approximately at 6 ms, the 1/1 HC significantly diminishes as the amplitude of the 3/1 XK approaches zero, leading to a weakened flux pumping process. Given that a sufficiently strong 1/1 HC is essential for initiating effective flux pumping to suppress the sawtooth, its weakened state results in inadequate suppression. Consequently, stronger sawtooth activity emerges.

\section{Conclusions and future perspectives}
We present a detailed investigation into the role of flux pumping mechanisms in the suppression of sawtooth oscillations in HBT-EP plasmas. Experiments show a clear correlation between the amplitude of edge modes and the emergence of strong sawtooth oscillations in HBT-EP plasmas. In HBT-EP, effective stabilization of the edge mode requires the normalized wall radius to be less than a critical value $(b/a)_c=$ 1.12. If this condition isn't met, sawtooth suppression occurs. During sawtooth-suppressed periods, our analysis identifies magnetic flux pumping as the likely mechanism responsible for sawtooth suppression. This process requires the coupling of several modes: the 1/1 helical core (HC), the 2/1 tearing mode (TM), and the 3/1 external kink mode (XK). Anomalous current broadening is also observed in these periods.

Central to the flux pumping mechanism is the 1/1 HC, driving the core dynamo effect. The concurrent existence of 3/1 XK, 2/1 TM and 1/1 HC suggests that the 1/1 HC could be influenced by components from both the 3/1 XK and the 2/1 TM. The interaction between the 3/1 XK and the conducting wall, which leads to the suppression of all three modes, results in a noticeable weakening, or possibly even termination, of the dynamo effect. There could be various scenarios: it might be that the 1/1 radial component from the 3/1 XK is the main contributor to the 1/1 HC, which then weakens as the 3/1 XK is stabilized by the wall. Alternatively, the 1/1 sideband from the 2/1 TM might be the primary source of the 1/1 HC, which also decays with the 3/1 XK since the 2/1 TM diminishes concurrently with the 3/1 XK. In both instances, the result is insufficient suppression, paving the way for the occurrence of pronounced sawtooth activities.

As we expand our understanding of sawtooth suppression mechanisms in HBT-EP plasmas, we recognize that certain aspects require further exploration. Although it is hypothesized that the 2/1 TM and the 3/1 XK both may contribute to the formation of the 1/1 HC and to the magnetic flux transport within the plasma, clarifying the role of each mode necessitates a methodology that can selectively dampen a mode without substantially altering the plasma profiles, a task that proves challenging with existing tools in HBT-EP, such as the bias probes, due to their pronounced impact on plasma conditions. Therefore, developing an approach to independently examine the effects of the 2/1 TM and the 3/1 XK is crucial. Achieving this would significantly enhance our comprehension of the mechanisms behind sawtooth suppression. 

In conclusion, our experimental observations are consistent with the requirements for magnetic flux pumping, and provide compelling evidence for its role in sawtooth suppression. In particular, we demonstrate how the stabilization of the edge mode leads to strong sawteeth. Sawtooth suppression by flux pumping does not depend on the source of the HC, and can occur in various beta regimes \cite{Bando21PPCF}. This means various sources to induce 1/1 structure in the core are possibly useful to realize sawtooth control via the dynamo effect, such as 2/1 NTM \cite{Bando21PPCF} and externally induced 3D field \cite{Piovesan17}. This study demonstrates the possibility to suppress sawtooth via an external mode, and suggests promising avenues for future research, including more precise sawtooth control through external mode manipulation.

\ack
This work was supported by U.S. Department of Energy, Office of Science, Office of Fusion Energy Science, Grant DE-FG02-86ER53222.

\appendix
\section*{Appendix}
To further illustrate how the normalized wall radius $b/a$ correlates with sawtooth behavior bifurcation, additional example discharges are presented. In the case of strongly sawtoothing discharges, it is observed that $b/a$ falls below the critical value $(b/a)_c=1.12$, thereby enabling the stabilization of the edge mode.

Conversely, for discharges where sawtooth activity is suppressed, the minimum $b/a$ still remains above the critical value $(b/a)_c$, and the strong edge mode continues to exist. This difference in the $b/a$ values relative to the critical value provides insight into the mechanisms influencing edge mode stabilization and sawtooth behavior in these discharges.

\begin{figure}
  \begin{minipage}[c]{0.45\textwidth}
    \subfloat{%
      \includegraphics[width=\linewidth]{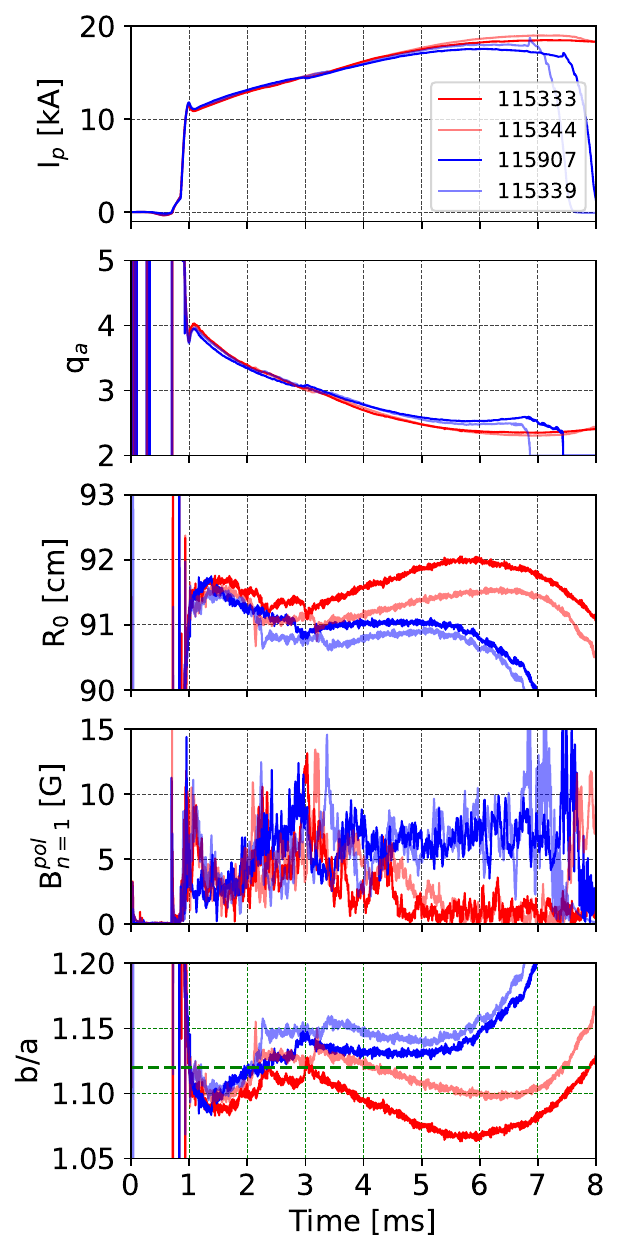}}
  \end{minipage}
  \hfill
  \begin{minipage}[c]{0.4\textwidth}
    \vspace{0pt} 
    \begin{minipage}[t]{\linewidth}
    \subfloat{%
        \includegraphics[width=\linewidth]{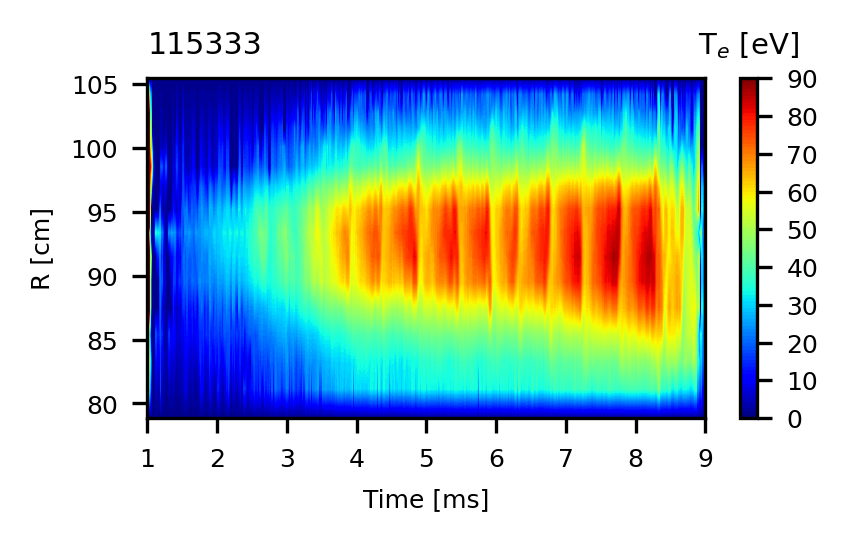}}
    \end{minipage}
    \\[0ex] 
    \begin{minipage}[b]{\linewidth}
    \subfloat{%
        \includegraphics[width=\linewidth]{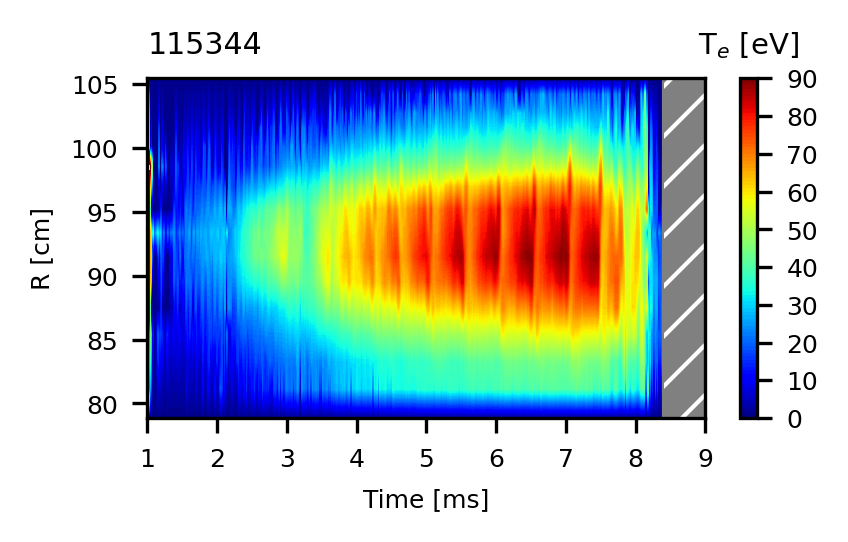}}
    \end{minipage}
    \\[0ex] 
    \begin{minipage}[b]{\linewidth}
    \subfloat{%
        \includegraphics[width=\linewidth]{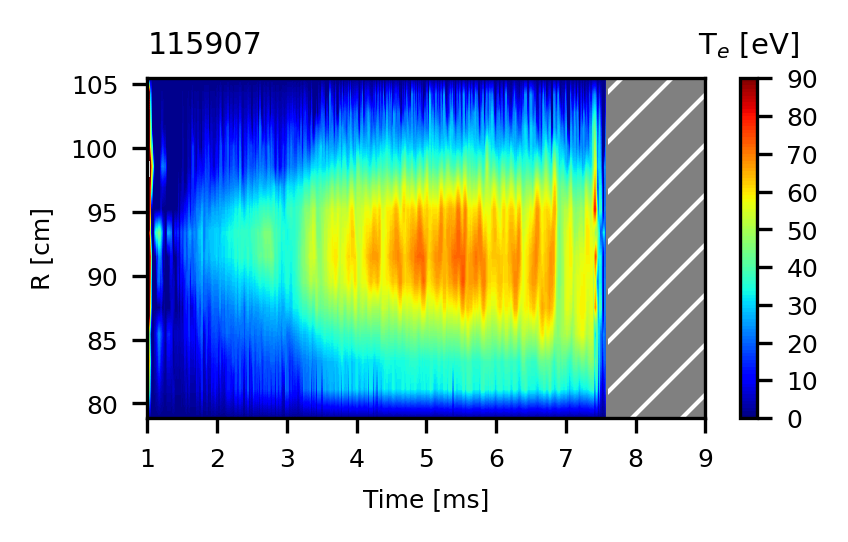}}
    \end{minipage}
    \\[0ex] 
    \begin{minipage}[b]{\linewidth}
    \subfloat{%
        \includegraphics[width=\linewidth]{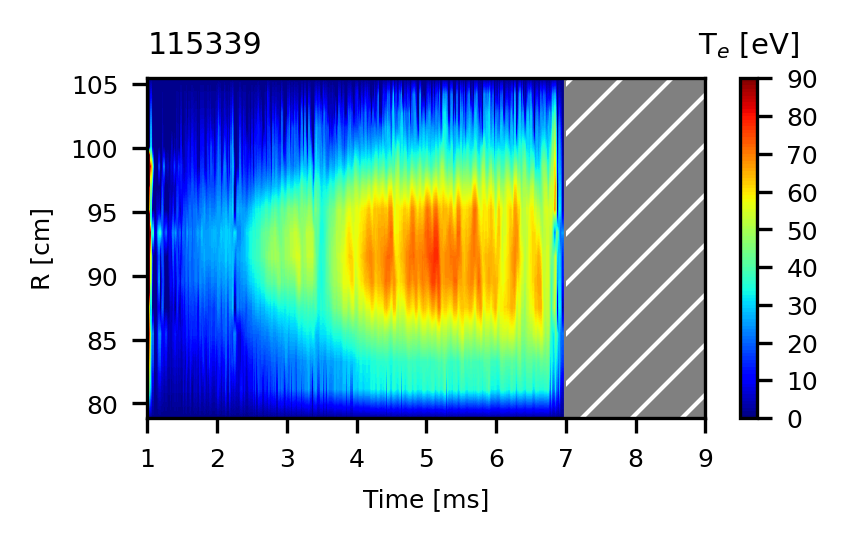}}
    \end{minipage}
  \end{minipage}
  \caption{More example discharges showing the bifurcation of sawtooth behaviors. The critical normalized wall radius $(b/a)_c=1.12$ is marked by the green dashed line.}
  \label{fig:ba}
\end{figure}

\section*{References}
\bibliography{IOPLaTeX}
\end{document}